\documentstyle[aps,preprint,epsfig]{revtex}
\begin{document}
\draft
\title{Polarons in Carbon Nanotubes}
\author{M. Verissimo-Alves$^1$, R. B. Capaz$^1$, Belita Koiller$^1$, Emilio 
Artacho$^2$, and H. Chacham$^3$}
\address{$^1$Instituto de F\'\i sica, Universidade Federal do Rio de Janeiro, 
Caixa Postal 68528, Rio de Janeiro, RJ, Brazil, 21945-970\\
$^2$Departamento de F\'\i sica de la Mat\'eria Condensada, C-III and Instituto 
Nicol\'as Cabrera, Universidad Aut\'onoma de Madrid, 28049 Madrid, Spain \\
$^3$Departamento de F\'\i sica, ICEx, Universidade Federal de Minas Gerais, 
Caixa Postal 702, 30123-970 Belo Horizonte, MG, Brazil}

\date{\today}
\maketitle
\begin{abstract}
We use {\it ab initio} total-energy calculations to predict the existence of 
polarons in semiconducting carbon nanotubes (CNTs). We find that the CNTs' band 
edge energies vary linearly and the elastic energy increases quadratically with 
both radial and with axial distortions, leading to the spontaneous formation of 
polarons. Using a continuum model parametrized by the {\it ab initio} 
calculations, we estimate electron and hole polaron lengths, energies and 
effective masses and analyze their complex dependence on CNT geometry. 
Implications of polaron effects on recently observed electro- and 
opto-mechanical behavior of CNTs are discussed. 
\end{abstract}

\pacs{PACS numbers: 71.38.+i, 71.20.Tx, 71.15.Mb}

Carbon nanotubes (CNTs) have recently attracted a great deal of interest 
for their unusual electronic and mechanical properties \cite{book_Dresselhaus}. 
In this context, the interplay between mechanical distortions and electronic
structure plays a central role. Mechanical distortions can be generally 
classified as externally applied or spontaneous. Effects of externally applied 
distortions such as twisting, bending and axial compression of CNTs on their 
electronic structure have been the subject of many studies 
\cite{Mazzoni_Chacham,Yang_Anantram,Park_Chang,Mazzoni1}. Spontaneous 
distortions are usually related to strong electron-phonon interactions, and 
classic examples are polaron formation in ionic solids 
\cite{Appel}, Peierls distortion in 1D metals \cite{Peierls} and related 
excitations (solitons and polarons) in conjugated polymers \cite{ssh_rev}. 
Symmetry-breaking distortions in fullerenes and CNTs have also been considered 
by several authors 
\cite{Lammert_Crespi,Harigaya1,mintmire,Dressel92,Harigaya2,chamon}. 

In this work we show that an extra electron or hole in a CNT causes a 
completely different kind of spontaneous distortion: a combined radial 
(breathing-mode-like) and axial distortion. This perturbation causes the band 
edge energies to vary linearly and the elastic energy to increase quadratically 
with the distortion parameters. Therefore, the total energy of the system has a 
minimum at a nonzero value of the distortion parameter and a polaron is formed. 

The distortions considered here are changes in the CNT radius and length, 
characterized by the radial and axial strain parameters 

\begin {equation}
\epsilon_r=\frac{(R-R_0)}{R_0} \mbox{ and } 
\epsilon_z=\frac{(\ell-\ell_0)}{\ell_0},
\label{distortion}
\end{equation}
respectively, where $R_0$ is the equilibrium radius for a neutral, undistorted 
CNT and $\ell_0$ is the equilibrium length of its unit cell. Consider a 
semiconducting CNT with a single extra electron at the bottom of the conduction 
band. Allowing $\epsilon_{r,z}$ to depend on $z$, the CNT axis direction, we 
write the change in total energy caused by this extra electron as 
\cite{Holstein_small}:

\begin{eqnarray}
\nonumber
E=&-&\frac{\hbar ^{2}}{2m_{eff}}\int_{-\infty }^{+\infty }\psi^{*}(z)
\frac{d^{2}}{dz^{2}}\psi (z)\;dz \\ \nonumber &+&\lambda_r\int_{-\infty }^
{+\infty }\psi ^{*}(z)\psi (z)\epsilon_r(z)\;dz \\ \nonumber &+&\lambda_z\int_
{-\infty }^{+\infty }\psi ^{*}(z)\psi (z)\epsilon_z(z)\;dz\;\\ \nonumber &+&\;
\frac{k_r}{2} \int_{-\infty }^{+\infty }\epsilon_r ^{2}(z)\;dz +\;\frac{k_z}{2}
\int_{-\infty }^{+\infty }\epsilon_z ^{2}(z)\;dz\;\\ &+&\;k_{rz} \int_{-\infty }
^{+\infty }\epsilon_r (z) \epsilon_z(z)\;dz ,
\label{totalenergy}
\end{eqnarray}
where $\psi(z)$ is the electronic wavefunction and $m_{eff}$ is the electron 
effective mass; $k_{r,z}$ and $\lambda_{r,z}$ are the effective spring 
constants per unit length and the electron-phonon coupling constants relative 
to the purely radial ($r$) and purely axial ($z$) strains; $k_{rz}$ is the 
spring constant relative to coupled radial and axial strains. Minimizing 
(\ref{totalenergy}) with respect to $\psi ^{*}$ and $\epsilon_{r,z}$ leads to 
the following expressions for the axial and radial strains:

\begin{equation}
\epsilon_{r,z}=\frac{\lambda_{z,r}k_{rz}-\lambda_{r,z}k_{z,r)}}
{k_rk_z-k_{rz}^2}\psi^*\psi=C_{r,z}\psi^*\psi
\label{distortions}
\end{equation}
and to the nonlinear Schr\"odinger equation
\begin{equation}
\left[ \frac{d^{2}}{dz^{2}}- \tilde C \psi^*(z)\psi(z) \right] 
\psi (z)= \tilde\varepsilon \;\psi (z)
\label{minimiza_psi}
\end{equation}
with $\tilde C=(2m_{eff}/\hbar^2)(\lambda_rC_r+\lambda_zC_z)$. Eq.
(\ref{minimiza_psi}) admits the following bound normalized solution:
\begin{equation}
\psi (z) = \sqrt{\frac{a}{2}} \; \mbox{sech}(az),
\label{psi}
\end{equation}
where the inverse polaron length, $a$, and its binding energy, 
$\varepsilon=-\hbar^2\tilde\varepsilon/2m_{eff}$, are given by

\begin{equation}
a=\frac{\tilde C}{4}\;\; ; \;\; 
\varepsilon=-\frac{\hbar^2} {2m_{eff}}\frac{\tilde C^2}{16}.
\label{final_coeffs}
\end{equation}
The resulting maximum axial and radial distortions are 
$\epsilon_{r,z}^{max}=aC_{r,z}/2$.

The polaron mass can be estimated using a semi-classical description of the 
electron motion. Assuming that the polaron propagates with velocity $v_{pol}$ 
without changing its characteristic shape, we write the total kinetic energy 
of the electron-lattice system as the sum of the energy of an electron 
propagating in the conduction band with velocity $v_{pol}$ plus the kinetic 
energy of the ions due to the propagation of the polaron:

\begin{equation}
T_{pol}=T_{e}+ T_{ions} = \frac{1}{2} m_{eff} v_{pol}^2+T_{ions} = 
\frac{1}{2} m_{pol} v_{pol}^2.
\label{total_kin}
\end{equation}
Both radial and axial components of the ionic velocities contribute to the 
ionic kinetic energy:
\begin{eqnarray}
\nonumber
T_{ions}&=&\frac{1}{2}\sum_{ions} M_{i}\left[R_0^2\left(\frac{\partial 
\epsilon_r}{\partial t}\right)^2+\ell_0^2\left(\frac{\partial \epsilon_z}
{\partial t}\right)^2\right]\\&=&\frac{1}{2}\sigma v_{pol}^2
\int^{+\infty}_{-\infty} dz\left[R_0^2\left(\frac{\partial \epsilon_r}
{\partial z}\right)^2+\ell_0^2\left(\frac{\partial \epsilon_z}{\partial z}
\right)^2\right]
\label{ion_kin_energy}
\end{eqnarray}
where $M_i$ is the ionic mass and $\sigma$ is the linear mass density of the 
CNT. Using Eqs. (\ref{distortions}) and (\ref{psi}), we obtain the polaron mass:

\begin{equation}
\frac{m_{pol}}{m_{eff}}=1+\frac{2\sigma a \tilde C^2}{15m_{eff}}\left[\left(
\frac{R_0C_r}{2}\right)^2+\left( \frac{\ell_0C_z}{2} \right)^2 \right]
\label{polaron_mass}
\end{equation}

Exactly the same results are obtained for an extra hole in the valence band, 
with a reversed sign in the definition of $\lambda$, as discussed below. All 
quantities defined above are functions only of $k_{r,z,rz}$, $\lambda_{r,z}$, 
$m_{eff}$, $R_0$, $\ell_0$ and $\sigma$. We determine $R_0$, $\ell_0$ and 
$\sigma$ from the CNT optimized geometry, and $m_{eff}$ is taken from tight-
binding (TB) calculations \cite{book_Dresselhaus} with a hopping parameter 
of $\gamma_0=-3.03$ eV. 

We obtain $k_{r,z,rz}$ and $\lambda_{r,z}$ from {\it ab initio} calculations. 
Our calculations are performed using the SIESTA\cite{siesta1} code, a 
numerical-atomic-orbital method based on density functional theory. This 
technique has been successfully applied to a number of studies involving 
nanotubes \cite{Mazzoni1,emilio}. We use a generalized gradient approximation 
for exchange and correlation and norm conserving pseudopotentials. A 
split-valence double-$\zeta$ basis of pseudoatomic orbitals with an orbital 
confining energy of 0.3 eV and an energy cutoff of 300 Ry for the fast Fourier 
transform integration mesh are used. In order to probe the dependence of 
polaron properties on CNT geometry, we perform calculations for (11,0) and 
(7,0) zigzag nanotubes. These are within the range of recently reported 
experimental values for nanotube diameters \cite{tang}. The k-point sampling 
is composed of six k-points in the CNT axis direction, which allows us to use 
a minimal hexagonal supercell of 44 atoms for a (11,0) CNT and 28 atoms for a 
(7,0) CNT.  The in-plane lattice parameter was chosen to be large enough (30 
\AA) to ensure that there is negligible interaction between periodic CNT images.

The calculations are performed for uniformly distorted neutral CNTs. Then 
$\epsilon_r$ and $\epsilon_z$ become independent of $z$ and the change in total 
energy per unit cell can be written as

\begin{equation}
\Delta E_{tot}=\ell_0\left(\frac{1}{2}k_r\epsilon_r^2+\frac{1}{2}k_z
\epsilon_z^2+k_rz\epsilon_r\epsilon_z\right).
\end{equation}

\begin{figure}
\epsfxsize=5.5cm
\epsfysize=7.5cm
\centerline{\epsfig{file=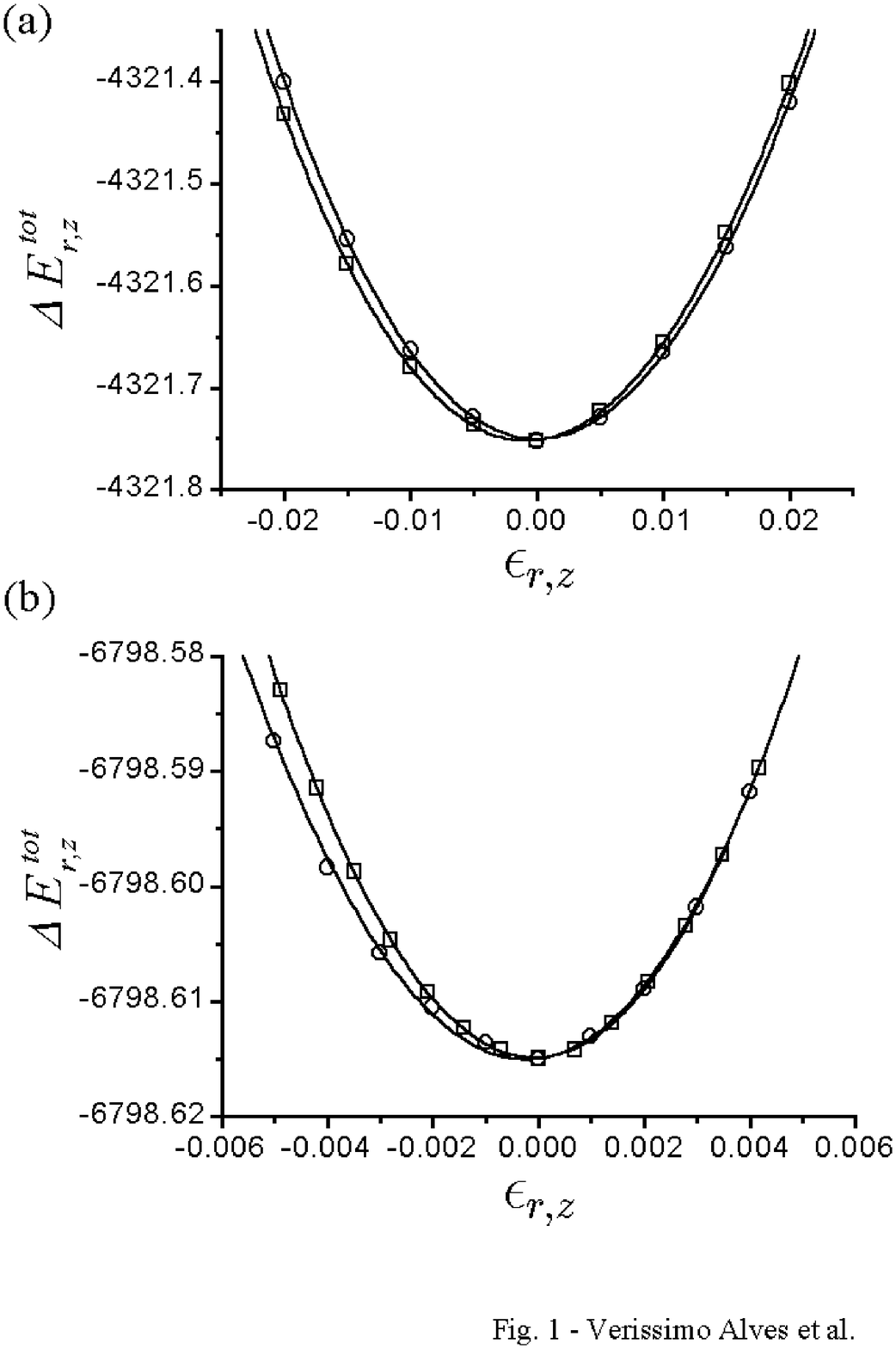}} 
\caption{Variation in total energy as a function of $\epsilon_r$ (triangles) 
and $\epsilon_z$ (squares) for (a) the (7,0) and (b) the (11,0) CNTs. The 
solid curve is a $3^{rd}$-order polynomial fit to the data. The values of 
$k_{r,z,rz}$ are obtained from the $2^{nd}$ derivative of the polynomial at the 
minimum.}
\label{fig1} 
\end{figure}

We vary the CNT radius while keeping $\epsilon_z=0$, and we vary the CNT unit 
cell length while keeping $\epsilon_r=0$: $k_r$ and $k_z$ are obtained, 
respectively, from the curvature on the $\Delta E_{tot}$ vs. $\epsilon_{r,z}$ 
plots, as shown in Fig. \ref{fig1}(a) for the (7,0) CNT and in Fig. \ref{fig1}
(b) for the (11,0) CNT. Then, by relaxing the unit cell length for a given 
nonzero value of the radial strain $\epsilon_r$, $k_{rz}$ is obtained. The 
resulting values for $k_r$, $k_z$ and $k_{rz}$ for both (11,0) and (7,0) 
CNTs are shown in Table \ref{table1}.

\begin{figure}
\epsfxsize=5.5cm
\epsfysize=7.5cm
\centerline{\epsfig{file=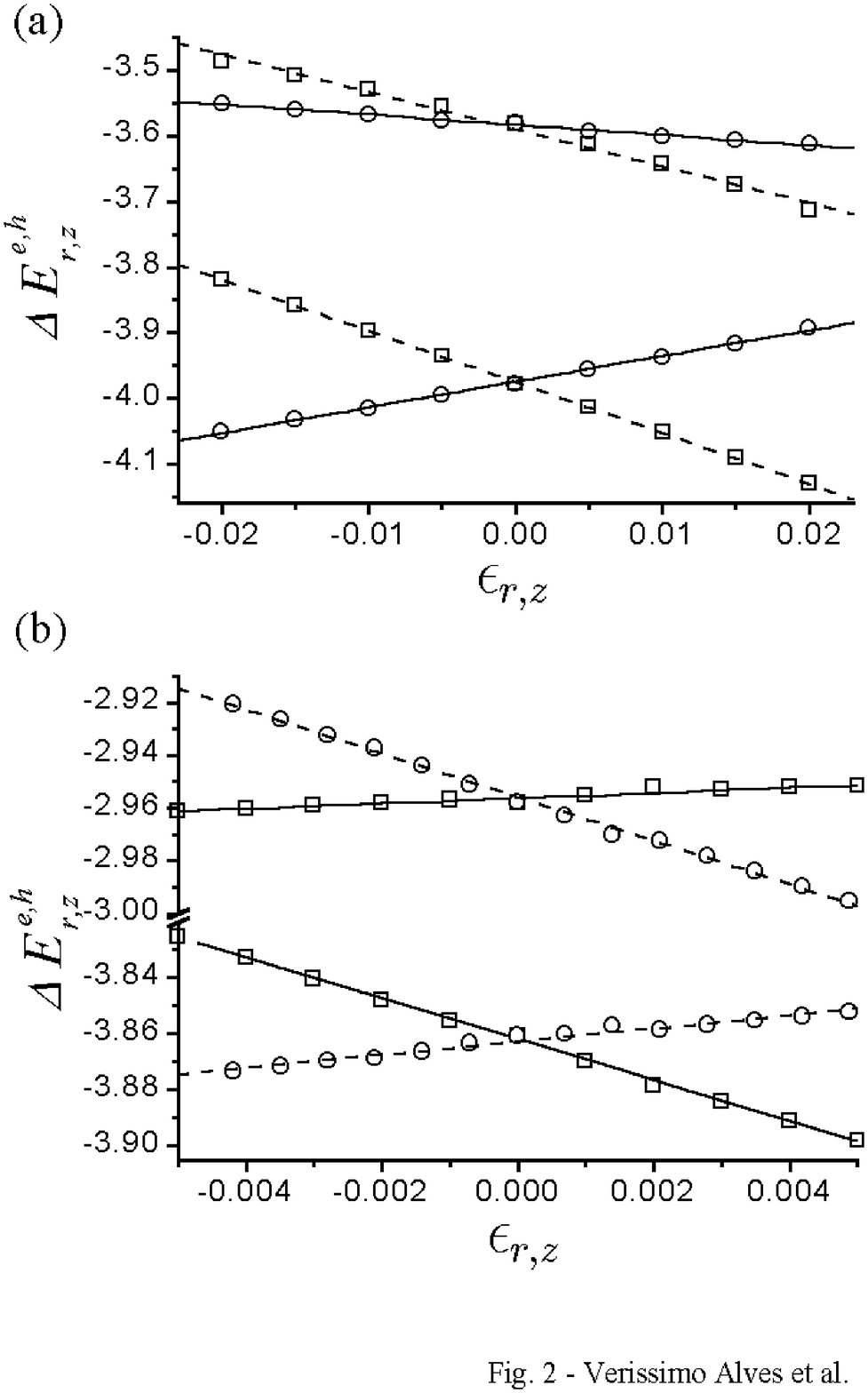}}
\caption{Band edge energies as a function of $\epsilon_r$ (squares) and 
$\epsilon_z$ (circles) for (a) the (7,0) and (b) the (11,0) CNT. Data around 
the regions $\epsilon_{r,z} \simeq 0$ are well fitted by straight lines.}
\label{fig2} 
\end{figure}

The electron-phonon coupling constants for electrons (holes), 
$\lambda^{e(h)}_{r,z}$, are obtained from the linear variation in the 
conduction (valence) band edge energies for a given strain energy 
$\epsilon_{r,z}$:

\begin{equation}
\Delta E_{r,z}^{e(h)}=\pm \lambda_{r,z}^{e(h)} \epsilon_{r,z},
\label{edge}
\end{equation}
where the positive sign is for electrons and the negative sign is for holes. 
Plots for $\Delta E_{r,z}^{e(h)}$ for both (11,0) and (7,0) CNTs are shown in 
Figs. \ref{fig2}(a) and (b), respectively. Values for all electron-phonon 
coupling constants are displayed in Table \ref{table1}.

The calculated polaron signatures (lengths, binding energies, masses and 
maximum distortions) calculated from the {\it ab initio} elastic and 
electron-phonon constants are also presented in Table \ref{table1}. The 
differences between the results for the (11,0) and (7,0) tubes allow us to 
anticipate a rich dependence on the CNT's diameter and chirality. This 
dependence can be understood as a superposition of two contributions: a 
``classical'' and a ``quantum'' contribution. The ``classical'' contribution 
comes simply from the dependence of the elastic constants on the CNT's 
diameter. From Table \ref{table1}, one sees that all elastic constants are 
smaller for the thinner (7,0) CNT. This effect can be understood in simple 
terms by considering CNTs with different diameters where the same percentual 
radial distortion is applied. Each bond in the CNTs' zigzag chains will be 
deformed by the same amount, but the thinner CNTs have less bonds along the 
chain. Considering the elastic energy to be proportional to the number of 
deformed bonds, the thinner CNTs will have smaller $k$'s. A similar argument 
applies to the axial elastic constants.  So, considering this ``classical'' 
argument alone, one would expect stronger polaron signatures for thinner CNTs. 

The ``quantum'' contribution comes from the different possible signs and 
magnitudes of the electron-phonon coupling constants $\lambda^{e(h)}_{r,z}$. 
As one sees from Table I, it is difficult to identify simple trends of the 
$\lambda$'s with geometry. Simpler TB models predict a constant value of 
$\lambda$ for zig-zag tubes under uniaxial strain \cite{Yang_Anantram}. This 
is probably an effect of curvature and rehybridization of $\sigma-\pi$ orbitals 
\cite{Blase} in small-diameter CNTs, which are generally neglected in TB 
calculations.

Analysis of polaron signatures in Table \ref{table1} might suggest that the 
observation of polarons in CNTs would be very difficult. Although we have not 
attempted to reach particular combinations of chirality and carrier type to 
achieve the strongest polaron signatures, the studied examples yield binding 
energies at most $\simeq 0.3$ meV, and the largest polaron mass $\simeq 0.2\%$ 
larger than the free electron mass for the hole polaron in the (7,0) CNT. 

However, these are signatures for a {\it single} polaron. The situation changes 
completely when we consider {\it collective} polaron effects. One example is 
the CNT length variation due to polarons. The total length variation on an CNT 
caused by the presence of a single polaron is:

\begin{equation}
\Delta \ell^{(1)}=\int _{-\infty} ^{\infty}dz\; \epsilon_z(z) = C_z
\label{ell} 
\end{equation}
For the hole polaron in the (7,0) CNT, $C_z=0.02$ \AA. Therefore, a modest 
number of 500 polarons would cause a sizeable 10 \AA$\;$ variation in the 
nanotube length, large enough to be observed, for instance, in AFM or STM 
experiments where CNTs are used as probes. A related quantity is the strain-
charge coefficient (SCC)\cite{Baughman},
\begin{equation}
\frac{\Delta \ell / \ell}{\Delta y}=\frac{N\Delta \ell^{(1)}}{\ell_0}= 
\frac{NC_z}{\ell_0},
\label{scc} 
\end{equation}
where $\Delta \ell / \ell$ is the fractional change in length caused by a 
$\Delta y$ change in the concentration of injected charge per carbon atom and 
$N$ is the number of atoms in the unit cell. Our calculated values for the SCC 
are also presented in Table \ref{table1}. They are comparable to the 
experimentally measured values of 0.17 (for low charge injection) in the 
context of electro-mechanical actuation in CNTs \cite{Baughman}. This 
suggests that polaron formation may contribute significantly to the observed 
actuation, although it may not be the only driving mechanism, since 
experiments are usually undertaken in much more complex environments where 
inter-tube interactions and other collective effects may be important.

Opto-mechanical effects in CNTs have also been observed \cite{iijima}. Our 
results allow us to predict strong or weak axial distortions as an elastic 
response to light, depending on the signs of $C_z$ for electrons and holes. 
Consider, for instance, electron-hole pairs generated by light  in (11,0) and 
(7,0) CNTs. From the signs of $C_z$ (the same signs of $\epsilon_z^{max}$ in 
Table \ref{table1}), we can see that an electron polaron causes an axial 
expansion in both (11,0) and (7,0) CNTs. On the other hand, a hole polaron 
causes an expansion in the (11,0) and a contraction in the (7,0). Therefore, 
for the (11,0) CNT, the axial mechanical effects of the electron and hole will 
add up and the CNT will have a strong elastic response to light. On the other 
hand, electron and hole strains will partly cancel out in the (7,0) CNT, 
leading to weaker opto-mechanical effects.

In conclusion, we predict the existence of polarons in semiconducting CNTs. 
Polaron properties are estimated from {\it ab initio} total-energy 
calculations for neutral uniformly distorted CNTs within a continuum 
approach. The complex dependence of polaron signatures on CNT chirality is 
understood as a combination of classical (elastic) and quantum effects. We 
show that collective polaron effects may have implications in the context of 
recently observed electro-mechanical \cite{Baughman} and opto-mechanical 
\cite{iijima} activities in CNTs. In particular, we predict the existence of 
two types of nanotubes regarding their elastic response (strong or weak) to 
light. A quantitative description of these effects should involve inclusion 
of collective behavior, inter-tube interactions and electrostatic interactions 
between electrons and holes should be taken into account into the model 
calculations.

\acknowledgements
We acknowledge fruitful discussions with M. S. C. Mazzoni. This work is 
partially supported by Brazilian agencies Conselho Nacional de 
Desenvolvimento Cient\'\i fico e Tecnol\'ogico (CNPq), Funda\c c\~ao 
Universit\'aria Jos\'e Bonif\'acio (FUJB), Programa de N\'ucleos de
Excel\^encia (PRONEX-MCT), Coordena\c c\~ao de Aperfei\c coamento de Pessoal 
de N\'\i vel Superior (CAPES) and Funda\c c\~ao de Amparo \`a Pesquisa do Rio 
de Janeiro (FAPERJ).

\begin{table}[htbp]
    \caption{Calculated quantities for a polaron in semiconducting zigzag CNTs.}
    \label{table1}
      \begin{tabular}{|c|c|c|c|c|}
      {\em Parameter} & \multicolumn{2}{c|}{(11,0) CNT} & 
\multicolumn{2}{c}{(7,0) CNT} \vline \\ \hline
          & electron & hole & electron & hole \\ \hline
        $\lambda_r$ (eV) & $0.99$ & $7.30$ & $-5.65$ & $-3.91$ \\ \hline
        $\lambda_z$ (eV) & $-8.29$ & $-2.38$ & $-1.55$ & $7.76$ \\ \hline
        $k_r$ (eV/\AA) & 605 & 605 & 194 & 194 \\ \hline
        $k_z$ (eV/\AA) & 644 & 644 & 198 & 198 \\ \hline
        $k_{rz}$ (eV/\AA) & 142 & 142 & 78 & 78 \\ \hline
        $\varepsilon$ (meV) & -7.7 $\times$ 10$^{-2}$ & -7.1 
$\times$ 10$^{-2}$ & -3.5$\times$10$^{-2}$ & -2.7 $\times$10$^{-1}$ \\ \hline
        $L$ (\AA) & 390 & 404 & 586 & 209 \\ \hline
        $m_{pol}/m_{eff}$ & 1.00020 & 1.00012 & 1.00001 & 1.00221 \\ \hline
        $\epsilon_r^{(max)}$ & $-6.291 \times 10^{-6}$ & $-1.684 \times 
10^{-5}$ & $1.205 \times 10^{-5}$ & $3.413 \times 10^{-5}$ \\ \hline
        $\epsilon_z^{(max)}$ & $1.783 \times 10^{-5}$ & $8.212 \times 10^{-6}$ 
& $9.623 \times 10^{-7}$ & $ -5.275 \times 10^{-5}$ \\ \hline
        $|\frac{\Delta \ell / \ell}{\Delta y}|$ & 0.142 & 0.068 & 0.074 & 
0.146 \\ 
      \end{tabular}
\end{table}

\end{document}